# Comparing the route-choice behavior of pedestrians around obstacles in a virtual experiment and a field study

Hongliu Li[1], Jun Zhang[1]*, Long Xia[1], Weiguo Song[1], Nikolai W.F. Bode[2]

[1]State Key Laboratory of Fire Science, University of Science and Technology of China, Jinzhai Road 96, Hefei, Anhui, P. R. China

[2]Department of Engineering Mathematics, University of Bristol, Bristol BS8 1UB, UK

# Highlights

- We analyze pedestrian route choice behavior around obstacles in a simple scenario
- We compare a field study and a virtual experiment with identical setups
- We find qualitatively similar results in both settings
- This helps to validate virtual experiments as a useful methodology

# Abstract

Pedestrians often need to decide between different routes they can use to reach their intended destinations, both during emergencies and in their daily lives. This route-choice behavior is important in determining traffic management, evacuation efficiency and building design. Here, we use field observations and a virtual experiment to study the route choice behavior of pedestrians around obstacles delimiting exit routes and examine the influence of three factors, namely the local distance to route starting points and the pedestrian density and walking speeds along routes. Crucially, both field study and virtual experiment consider the same scenario which allows us to directly assess the validity of testing pedestrian behavior in virtual environments. We find that in both data sets the proportion of people who choose a closer exit route increases as the difference in distance between exit route starting points increases. Pedestrians' choices in our data also depend on pedestrian density along routes, with people preferring less used routes. Our results thus confirm previously established route choice mechanisms and we can predict over 74% of choices based on these factors. The qualitative agreement in results between the field study and the virtual experiment suggests that in simple route-choice scenarios, such as the one we investigate here, virtual experiments can be a valid experimental technique for studying pedestrian behavior. We therefore provide much-needed empirical support for the emerging paradigm of experiments in virtual environments.

* Corresponding author: Jun Zhang, Email: junz@ustc.edu.cn

## 1. Introduction

Pedestrians are a common sight in cities, transport hubs and at large public events like concerts and soccer games. A sad but recurring reminder for the importance of studying the movement of pedestrians are large-scale accidents, such as at the Love parade in Germany (Helbing and Mukerji, 2012) and at the Mecca pilgrimage (Alaska et al., 2017). In addition to understanding or even preventing such accidents, research into pedestrian dynamics promises to inform the design of buildings or pedestrian infrastructure and the implementation of traffic management approaches.

One crucial component of pedestrian behavior is the decisions individuals make on which route to use to reach their intended destinations. These decisions can cover different spatial scales. On the one hand, pedestrians may decide on routes over long distances, such as their commute to work, or a shopping or sightseeing trip. On the other hand, pedestrians also have to make decisions on much smaller spatial scales. For example, pedestrians can choose between two routes to walk around a freestanding obstacle (left or right in the direction of movement) and in confined indoor spaces, pedestrians can often choose between several exits. Here, we focus on the latter scenario and consider pedestrian choices on routes covering small spatial scales. These choices may be simple and not require any conscious decision-making process.

Previous work identifies a large collection of factors that can influence route choice even on the small spatial scales we are considering here. For example, factors that are considered to be influential include: the information about routes available to individuals, the desire for maintaining cohesion in groups of friends or families, visibility conditions, crowding of routes, social interactions including leadership by some individuals, the distance to exits and the directness of routes (Cao et al., 2018; Haghani and Sarvi, 2016, 2017a; Liao et al., 2017; Lovreglio et al., 2016; Miller et al., 2013; Shen et al., 2014; Srinivasan et al., 2017; Zhu and Shi, 2016). A different line of work focuses on more detailed aspects related to the motor control, steering, perception and cognition of individual pedestrians avoiding stationary obstacles (Fajen and Warren, 2003; Fink et al., 2007; Gérin-Lajoie et al., 2008; Gérin-Lajoie et al., 2005, 2006; Patla, 1997; Patla et al., 2004; Vallis and McFadyen, 2003, 2005). Instead of considering all of these aspects, we examine the relative influence of three different factors on pedestrian route choice: the distance to be covered, the congestion along routes and the walking speeds of other pedestrians using routes. We select these factors for two main reasons. First, previous work on these factors suggests how they affect route choice, as discussed below. This means they form an appropriate basis on which to perform a validation of different research methodologies, which is our primary contribution. Second, even under normal conditions and without experimental manipulation, a broad range of values describing the three factors can be expected to arise naturally in appropriately designed pedestrian facilities. This means a study, where the natural behavior of pedestrians is observed without experimental instructions, as employed by us here, is suitable for studying them.

Based on the literature, we expect the three factors we investigate to have the following effects on the route choice of pedestrians around obstacles. In general, under experimental conditions, pedestrians appear to prefer the route they expect to complete in the shortest time ("quickest path", e.g. (Kemloh Wagoum et al., 2012)). This means we can expect pedestrians to prefer shorter routes (Bovy and Stern, 2012; Guy, 1987; Liao et al., 2017; Seneviratne and Morrall, 1985; Verlander and Heydecker, 1997) and it is thus plausible to expect that people preferentially choose exits that are closer to them or shorter routes around obstacles. Similarly, this suggests that pedestrians avoid queues (Bode et al., 2014, 2015b; Bode and Codling, 2013; Haghani and Sarvi, 2016; Liao et al., 2017; Zhang and He, 2014) and we therefore expect that individuals select exits or routes around obstacles that are associated with a lower local density of pedestrians. Finally, time-dependent information, such as the speed at which other pedestrians move in the local neighborhood of exits or obstacles can be indicative of how quickly different routes can be completed and pedestrians may thus be drawn toward local areas of higher average walking speeds (Bode et al., 2015b)**.** We do not aim to uncover fundamentally novel aspects of pedestrian route choice behavior here, but a secondary contribution of our work is a confirmation of previous findings.

The main contribution and novelty of our work is a direct comparison of pedestrian route choice behavior based on the three factors discussed above between a natural setting and a virtual environment setting. Virtual experiments form a promising paradigm that can offer real-time visualization of a wide variety of information at different levels of immersiveness. Virtual environments are gaining popularity as a complementary experimental framework, as they are flexible, allow a high level of experimental control, can be conducted remotely and allow simulating high-pressure situations that could be dangerous if conducted with many volunteers, for example (Andrée et al., 2016; Bode et al., 2014, 2015b; Bode and Codling, 2013; Deb et al., 2017; Hartmann, 2010; Kinateder et al., 2014a; Kinateder et al., 2014b; Moussaid et al., 2016; Ronchi et al., 2015). To give examples of the scenarios that this emergent technology can be used for, consider (Chu et al., 2017) which calibrates a logistic regression model for guidance compliance behavior based on a virtual reality experiment. (Kinateder et al., 2014b) studies the impact of social influence on route choice behavior of pedestrians in a virtual reality tunnel fire. (Ye et al., 2018) examines day-to-day route choice models based on a virtual route choice experiment. Despite their evident popularity, virtual experiments face one fundamental problem: it is not clear to what extent the behavior observed in a virtual environment can be extrapolated to the real world (Kinateder et al., 2014c). In other words, the ecological validity of virtual experiments is often not established. Some studies have started to compare real-life pedestrian behavior to the behavior observed in virtual environments. For example, several studies investigate whether virtual experiments can be used to capture the road-crossing behavior of pedestrians in the presence of vehicles and generally find a good match to real-life behavior (Bhagavathula et al., 2018; Deb et al., 2017; Schwebel et al., 2008). Other research confirms that simple avoidance maneuvers between pedestrians can also be faithfully captured in virtual experiments (Iryo-Asano et al., 2018; Moussaid et al., 2016). One study that finds a good match in avoidance behavior between pedestrians in a real-life and a virtual experimental setting also finds that measures capturing the flow of pedestrians through a bottleneck cannot be reproduced accurately in the virtual environment (Moussaid et al., 2016).

This research suggests that virtual experiments are capable of faithfully reproducing pedestrian behavior. However, a confirmation that choices of pedestrians on available routes based on multiple factors, such as the ones we discuss above, can be reproduced in virtual experiments is missing to date. In addition, validation studies for virtual experiments have so far focused on highly immersive virtual environments that attempt to mimic the visual perception of pedestrians in three dimensions. However, more abstract virtual experiments in which participants have a top-down view on a virtual environment have been proposed as a plausible research tool but remain unvalidated to date (Bode et al., 2014, 2015b; Bode and Codling, 2013). The only study that starts to address these questions compares route choices in controlled experiments with data from interviews in which participants stated their preferred routes for a static top-down snapshot of a building floorplan that indicates the locations of other people (Haghani and Sarvi, 2017b). The authors find that factors, such as distance to exits and congestion, have similar effects on route choices in both data sets. In contrast to this approach of investigating stated exit choices based on static information, we consider dynamically varying virtual environments that participants can interact with and compare human route choice behavior in a highly abstracted virtual experiment (unclear ecological validity) to field observations (highest ecological validity).

To summarize, the two research questions of our work are as follows:

1. Can we confirm the effects of three factors (local distance to routes and the pedestrian density and walking speeds along routes) on route choice of pedestrians as predicted by the literature (nature of effects is discussed above)? This will confirm previous findings in a novel context.

2. Are these effects qualitatively the same in observational data of real pedestrian behavior and in a virtual experiment with a nearly identical route choice setup? This will provide empirical evidence on the similarity of human route choice behavior in real pedestrians and of human participants in simple virtual experiments on pedestrian behavior, an emerging technology that is rapidly gaining popularity.

The remainder of this article is structured as follows. In section 2, we describe the data collection procedure for the field study and the virtual experiment. In section 3, we present the results from both studies including our statistical analysis of the data and in section 4 we discuss our findings and draw conclusions.

## 2. Methods

We focus on the effects of three factors (pedestrian density and speed along routes and the distance to different route starting points) on pedestrian route choice around obstacles. The layout of the pedestrian facilities used for our data collection is designed in such a way that individual pedestrians must choose between several discrete routes around obstacles to reach their final goal. The usage and length of routes differ as a result of the movement choices of other pedestrians and of where pedestrians enter the facility. This means that measures related to the three factors we study vary naturally, allowing us to investigate their effect on pedestrian route choices in an observational setting without instructions to participants. The virtual experiment closely emulates this set-up and we deliberately choose a highly abstracted virtual environment and further justify this below.

Our data collection has the approval of the ethics committee of the University of Science and Technology of China. In the field observation, we record data from the movement of visitors at an exhibition with visitors being unaware of being filmed. We give hats to visitors prior to filming their movement from above to avoid recording identifiable facial features. In the virtual experiment, participants are informed about the nature of the experiment and consent to taking part via a tick box prior to starting the experiment. We record no identifying information in the experiment.

### 2.1 Field observation

Our field observation data is based on observations during a Sci-Tech exhibition at the University of Science and Technology of China in 2017, which lasted for two days and attracted more than 8000 visitors from a broad range of backgrounds including school classes, families and students. To study the movement of pedestrians, we use barriers to construct several different route choice scenarios directly in front of the door of an exhibition hall. After attending a science show inside the exhibition hall, visitors exit the hall by passing through our setup. Before entering the hall, visitors are given colored hats, which help to conceal identifying facial features and facilitate the image analysis involved in tracking the movement paths of pedestrians. Visitors are not informed about the reason for wearing a hat. We record the entire scenario from above using two HDR-SR11 cameras that are located on the fifth floor of the exhibition hall building (see https://doi.org/10.5281/zenodo.3365464 for the supplementary video and trajectories).

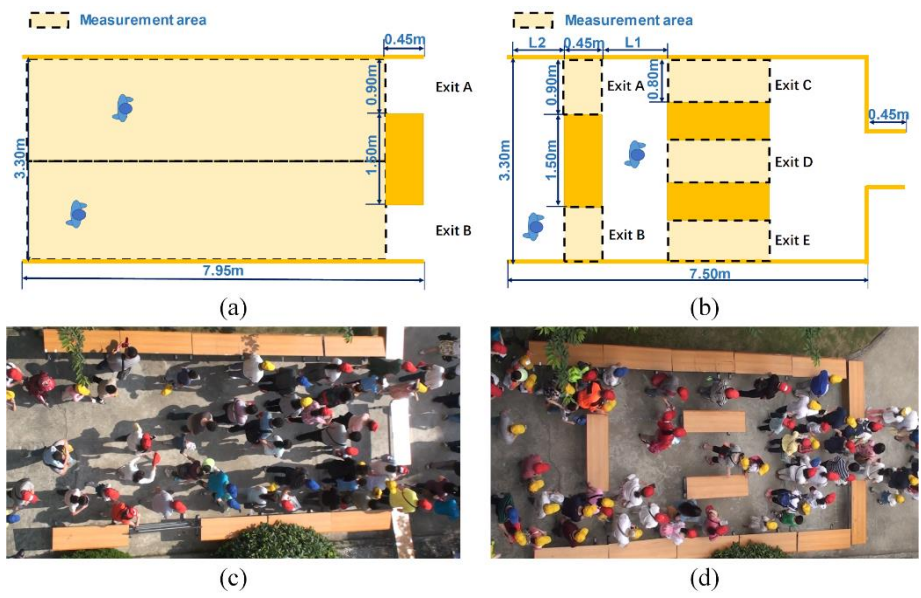

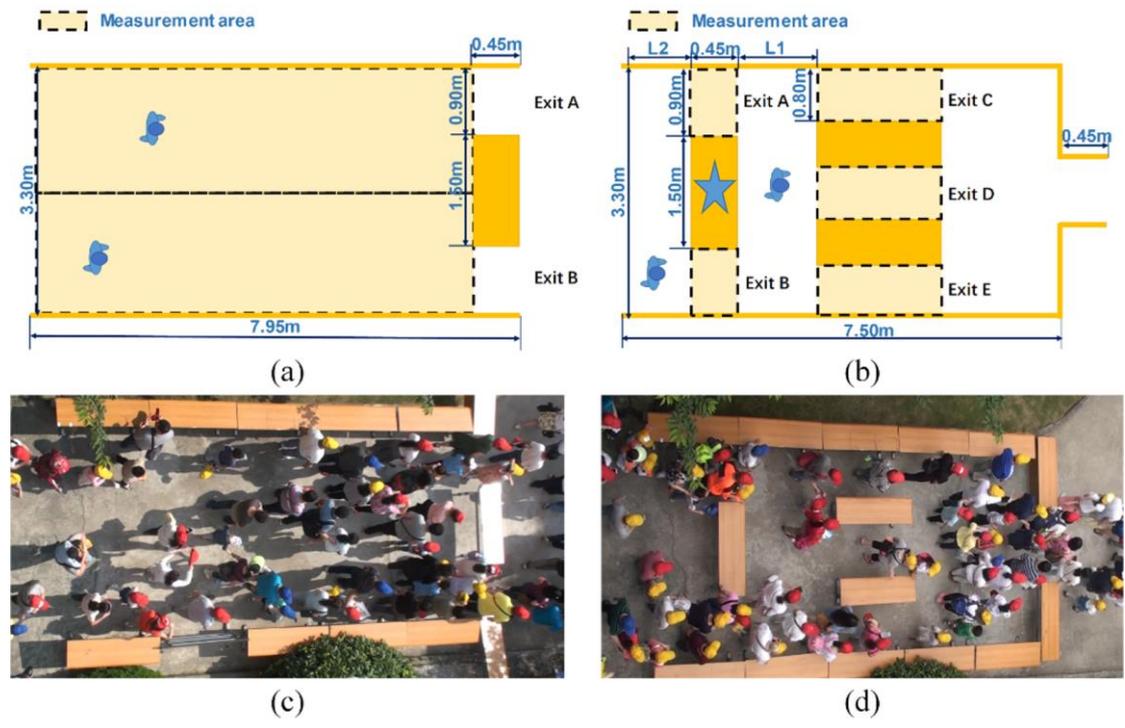

Fig 1. Plan of the scenarios we investigate in the field observation (a, b) and two still images from the study (c, d; measurement areas used in section 3). Orange lines and blocks in (a, b) indicate barriers. For the second scenario (b, d), we consider three experimental conditions obtained by moving the table marked with a blue star along the horizontal axis in panel (b). This results in different combinations of distances $L_1$ and $L_2$ between obstacles. We consider the following three ($L_1$, $L_2$) combinations: (1.75 m, 0.7 m), (1.65 m, 0.8 m), (1.55 m, 0.9 m). Pedestrians walk from the left of the scenario to the exit on the right.

We analyze data obtained from two types of symmetrical scenarios shown in Fig 1. The two scenarios differ in the number of routes pedestrians can choose from and thus capture a wide range of movement choices. The scene of the observation area is made up by tables (1.50 m × 0.45 m × 0.70 m) and measures 7.95 m × 3.30 m. In the first scenario type, pedestrians only have to choose once between two exit routes (Fig. 1 (a)). In contrast, in the second scenario type, pedestrians first have to choose between two exit routes, followed by a second choice between three exit routes (Fig. 1 (b)). In the second scenario type, we vary the distances between the tables outlining exit routes, as detailed in Fig. 1. In order to simplify the description, we name the first scenario (see Fig 1 (a), (c)) *G0* and the three different settings of the second scenario (see Fig 1 (b), (d)) *G1, G2* and *G3,* respectively. To facilitate comparison to future studies and to illustrate the demographic composition of our study population, we record information on the gender and social group composition of pedestrians (inferred from video recordings, see Appendix A).

In total, 233 pedestrians (44% females and 56% males) are recorded in the first scenario, G0, in 1.76 minutes, while 212 pedestrians (43.87% females and 56.13% males), 89 pedestrians (40.45% females and 59.55% males) and 164 pedestrians (34.15% females and 65.85% males) are recorded for scenarios G1, G2 and G3 in 2.47 min, 1.29 min and 1.74 min, respectively. We use the software *PeTrack* (Boltes et al., 2010) to obtain the trajectories (see Appendix B) of pedestrians from the video recordings. The trajectories of all pedestrians are recorded, regardless of whether they wear their hat or not. We manually check that all pedestrians are detected and tracked by the software.

## 2.2 Virtual experiment

Our virtual experiment on pedestrian route choice around obstacles is designed to present participants with a route choice scenario that closely matches the setup of the observational study. Key differences are that human participants have a top-down perspective and all pedestrians except the one controlled by participants are computer-simulated. Investigating an almost identical route choice scenario in the virtual experiment and the observational study allows us to directly compare the route choice behavior around obstacles of real pedestrians and humans controlling pedestrians in a virtual experiment.

The simulation framework for the virtual experiment extends established methodology (Bode et al., 2014; Bode and Codling, 2013). In total, we recruit 146 volunteers at the University of Science and Technology of China to take part in the experiment in 2018. Most of the participants are students. The age of participants ranges from 16 to 33 with an average of 22.74 years old. Since 8 participants accidentally stop the program before it finished, we only use data from 138 participants (60.14% males and 39.86% females).

In our experiment, participants have a top-down view of a virtual environment and control the movement of one pedestrian via mouse clicks in the presence of other, computer-simulated, pedestrians. The movement of simulated pedestrians follows a derivative of a previously developed model (Helbing et al., 2000). According to this model, pedestrians move in continuous two-dimensional space. Interactions between pedestrians (e.g. for collision avoidance) and between pedestrians and obstacles, such as walls, are implemented via forces that depend on the distance between pedestrians or between pedestrians and obstacles (Bode and Codling, 2013; Helbing et al., 2000).

The preferred movement direction of the participant-controlled pedestrian is given by the direction between the current location of this pedestrian and the most recent mouse click location of the participant. The preferred movement direction of computer-controlled pedestrians is given by the local gradient of a floor field that directs pedestrians to a final target. Details on the floor field and the simulations underlying the virtual environment can be found in (Bode et al., 2014; Bode and Codling, 2013). The floor plan of the virtual environment closely matches the two different obstacle avoidance scenarios we investigate in the field observation. It comprises of an entrance area, obstacles (black rectangles) and a target area (green rectangle; see Fig 2). We implement two scenarios S1 (Fig. 2 (a), matching G0) and S2 (Fig. 2 (b), matching G1-G3). The floor field used in our simulation model is implemented on a grid of 0.1×0.1 m cells (Bode and Codling, 2013) which means we can only represent G1-G3 in the virtual experiment up to an accuracy of 0.05 m. We decide to only implement one experimental condition, S2, in the virtual experiment to represent G1-G3, but we investigate the effect of different experimental conditions in detail (see below). Details of the experimental scene setup can be seen in Fig. 2 (note that S2 matches G1 up to an accuracy of 0.05 m).

Participants complete a total of 12 tasks, 6 each in S1 and S2, respectively. In each task, the simulated pedestrian controlled by the participant is placed at an initial position inside the starting area. Participants then have to move to a green target area via a route of their choice. To produce a range of different scenarios for participants to respond to, we vary the number, initial position and preferred movement routes of simulated pedestrians. Across tasks, we vary the initial position of participant-controlled pedestrians, choosing randomly between 2 different horizontal positions (to broaden the data range of the distance factor) and 7 equidistant vertical positions in the entrance area. Fig. 2 (c) and (d) shows two distinct vertical initial positions for the same horizontal position. The second horizontal position is shifted 1.5 m in simulated space to the right of the position shown in Fig. 2 (approximately 3 shoulder widths of simulated pedestrians). In this way, we ensure that participant-controlled pedestrians are initially closer to some routes than others. We also vary the number of computer-controlled pedestrians to change the pedestrian density participants experience. We simulate either 9 or 19 computer-controlled pedestrians. In addition, we alter the floor field directing the movement of computer-controlled pedestrians. For both S1 and S2, we implement three separate floor fields that encode different movement preferences. The floor fields implement a dividing horizontal line, above which computer-controlled pedestrians move along routes closer to the top of the screen display of the virtual environment and vice versa. The three floor fields place the dividing line either centrally in the vertical direction (symmetric preferences), or shifted downwards or upwards (i.e. the likelihood for randomly placed pedestrians to move along routes closer to the top increases or decreases, respectively). We refer to these movement preferences as 'symmetric', 'up' and 'down', respectively. Each participant is exposed to the same sequence of scenario, pedestrian number and pedestrian movement preference combinations across the 12 tasks, as follows: (S1, 9, symmetric); (S1, 19, up); (S2, 19, up); (S1, 9, up); (S2, 9, down); (S2, 9, symmetric); (S2, 19, down); (S1, 19, down); (S2, 9, up); (S1, 19, symmetric); (S1, 9, down); (S2, 19, symmetric). The initial positions of computer-controlled pedestrians are determined randomly (uniformly) within the two-thirds of the virtual environment further away from the target area. The effect of this implementation of tasks is that participants face a range of situations within S1 and S2 that differ with respect to participants' relative position to exit routes and exit route usage by computer-simulated pedestrian.

Each participant completes the experiment, including all 12 tasks, only once. Before the experiment, participants are given instructions on how to steer the pedestrian they control and on what they are expected to do. They are also told to imagine that they are leaving from an exhibition. Participants are not allowed to watch the experiment before taking part and participants who have already finished the experiment are not allowed to share their experience with others who have not yet taken part. During the experiment, we only answer questions on how to steer the pedestrian. We record the locations of all pedestrians in the virtual environment, the on-screen location of mouse clicks, as well as the number of mouse clicks and the time span to the first click for each participant (see https://doi.org/10.5281/zenodo.3365464 for the supplementary video and trajectories) .

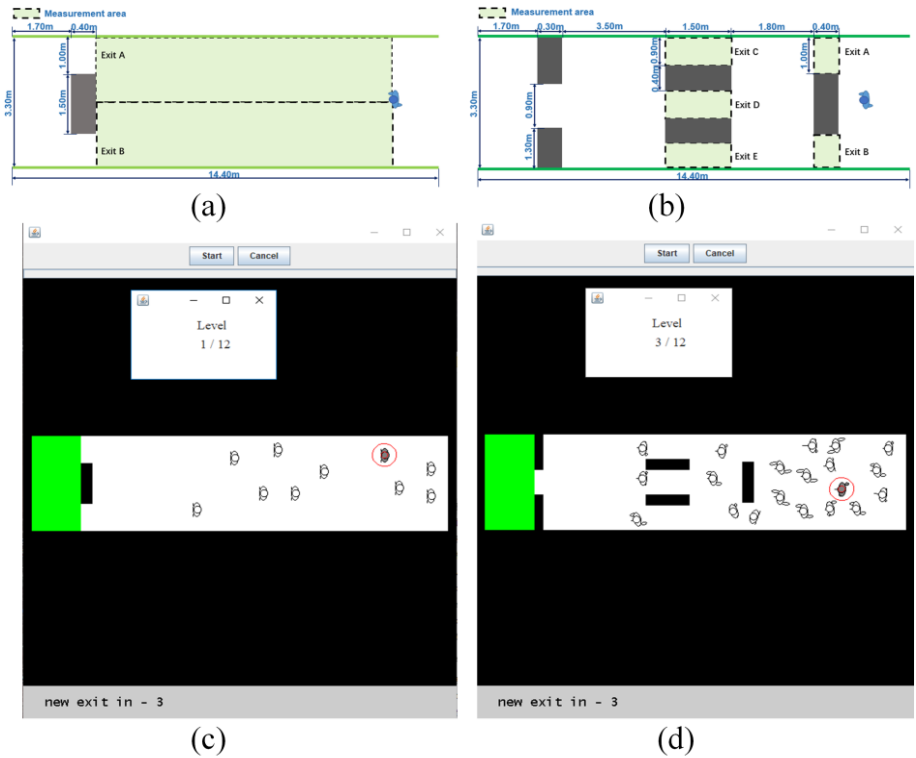

Fig 2. Sketch and screenshots of virtual experiment for scenarios S1 (a, c) and S2 (b, d) (4 pixels = 0.1m). In each scenario, there are six tasks (different initial positions and distributions of simulated agents) for each participant. Simulated pedestrians are represented by white filled circles and the pedestrian steered by participants is shown in grey and is highlighted with a red circle in the screenshots above. The black rectangles are obstacles. Participants control the grey agent via mouse clicks and move from their starting position (shown) to the target area (green region).

The top-down view participants have in our virtual environment differs substantially from how pedestrians perceive their environment in the real world. For our study, we deliberately use a highly abstracted set-up, which is situated at the lower end of the level of immersiveness achievable in virtual experiments. In this way, we test if the mechanisms by which people choose routes around obstacles are robust, even when individuals' perception and movement control are different. As the abstracted virtual experiment set-up we use here is proven to be flexible, easy to use and not dependent on specialist equipment, we suggest this comparison is particularly useful for the research community.

## 3. Data analysis and results

In this section, we present our data analysis and results for the virtual experiment and the field observation. We focus on the effects of three factors on the route choice of pedestrians around obstacles (the difference of density around two neighboring routes, the difference in average speeds of pedestrians using two adjacent routes and distance from the positions of pedestrians to different routes). The distance, density and speed are analyzed to study whether pedestrians prefer the closer, the less busy or the route used by pedestrians moving at higher speeds. Before presenting a formal statistical analysis of the data in section 3.4, we describe how we measure each of the three factors in turn and provide illustrative results for their effects in sections 3.1-3.3. The illustrative results are useful to indicate general trends in the data, but due to the possible influence of multiple factors on pedestrian behavior, they are not suited for a comparison between observational data and virtual experiments. Such a comparison can be performed based on the statistical analysis in section 3.4.

### 3.1 Distance to the exit routes

Previous work has shown that pedestrians prefer shorter routes (Liao et al., 2017) and it is thus plausible to expect that people preferentially choose exit routes that are closer to them. In order to study the impact of distance on pedestrians' route choice around obstacles, we calculate the Euclidean Distance $d$ from the position of individuals to the mid-point of the opening between obstacles or obstacles and walls ((from now on referred to as 'exits'). In scenarios G0 and S1, we measure $d$ from the initial position of each participant to the mid-points of exits A and B (see Fig. 1 (a)). The first choice in scenarios G1-G3 and S2 is treated in the same way and for the second choice, we measure $d$ from the location individuals are at when they have just left the measurement area marked in the figure for either exit A or B (see Fig 1 (b)).

We find that in both the field observation and the virtual experiment, pedestrians prefer to choose the nearer exit, as indicated by $d$. In the field observation, in G0, 82.9% participants choose the closer exit. However, only 56.47% participants choose the closer exit in the virtual experiment S1. Considering scenarios G1-G3 and S2, we can see that if a pedestrian initially chooses exit A (alternatively B), the route to the final exit is the same length whether he or she subsequently chooses Exit C or D (alternatively D or E). But when choosing Exit A (B) individuals are closer to Exit C (E) immediately after passing through the first exit. Our findings suggest that pedestrians choose their routes mainly based on local distance instead of global distance. For example, 91.04% pedestrians in G1, 98.88% in G2 and 95.73% in G3 choose shorter route in the first choice. Subsequently, 63.03%, 71.91% and 72.56% of them choose local nearest exit in the second choice in G1, G2 and G3, respectively. The virtual experiment produces similar trends with 73.16% and 89.73% participants choosing the local closer exit in the first choice and in the second choice in S2, respectively.

To illustrate this local route choice behavior, we divide the observation area for the field observation into a grid of 5 cm×5 cm cells and draw a frequency distribution map of pedestrians' positions based on trajectories for scenarios G1-G3 (Fig. 3). When we consider pedestrians who initially select exits A or B separately, the frequency of pedestrians using the local shorter route can be seen clearly (see left and middle columns in Fig 3). For example, in G1, more pedestrians who initially choose exit A subsequently follow the top route (using Exit C; top left-hand panel in Fig. 3).

Considering data regardless of initial choices (right-hand column in Fig. 3), in G1, the numbers of pedestrians choosing the Exit A and B are nearly the same and thus the fraction of pedestrians choosing Exit C, D and E is relatively balanced. In contrast, there is an overall bias towards Exit B in G2 and G3 and as a result, the fraction of pedestrian choosing Exit E is higher.

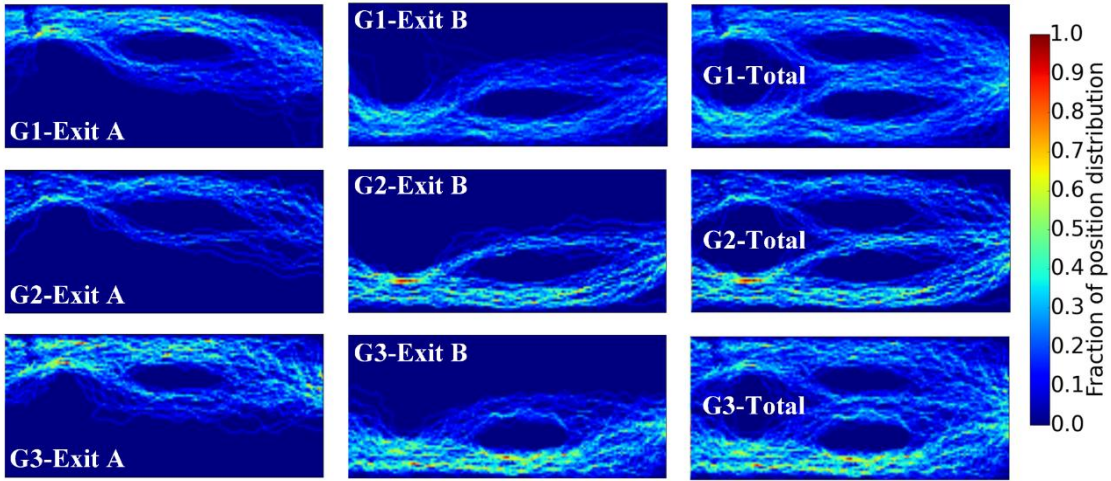

Fig 3. Frequency distribution map of pedestrians' positions in G1-3 in the field observation. The field observation space is divided into a grid of 5 cm×5 cm cells and the frequency of visits to each grid location is calculated. Colors indicate the normalized location frequency which is computed separately for each diagram. Rows show data from the scenarios G1-G3 and columns show the data for each scenario, considering only individuals who initially choose Exit A (see Fig. 1 (b)), individuals who initially choose Exit B and considering all individuals.

To show how the strength of the preference pedestrians have for nearer exits depends on the difference in distance between exits, we calculate the distance difference $\Delta d$ between the distances $d$ to two neighboring exits. For the second choice in scenarios G1-G3 and S2, we only consider the two closer exits and do not consider data for individuals who choose the exits furthest away (1 and 2 people in G1-G3 and S2, respectively). For consistency, we always subtract the shorter distance from the longer distance and plot this measure $\Delta d$ against the fraction of pedestrians selecting the closer exit (Fig. 4). We plot this relationship separately for first and second exit choices and find that the fraction of pedestrians choosing the closer exits increases with $\Delta d$ in all cases and for both the field observation and the virtual experiment (Fig. 4). When $\Delta d$ exceeds 1.0 m, nearly all pedestrians choose the closer exit.

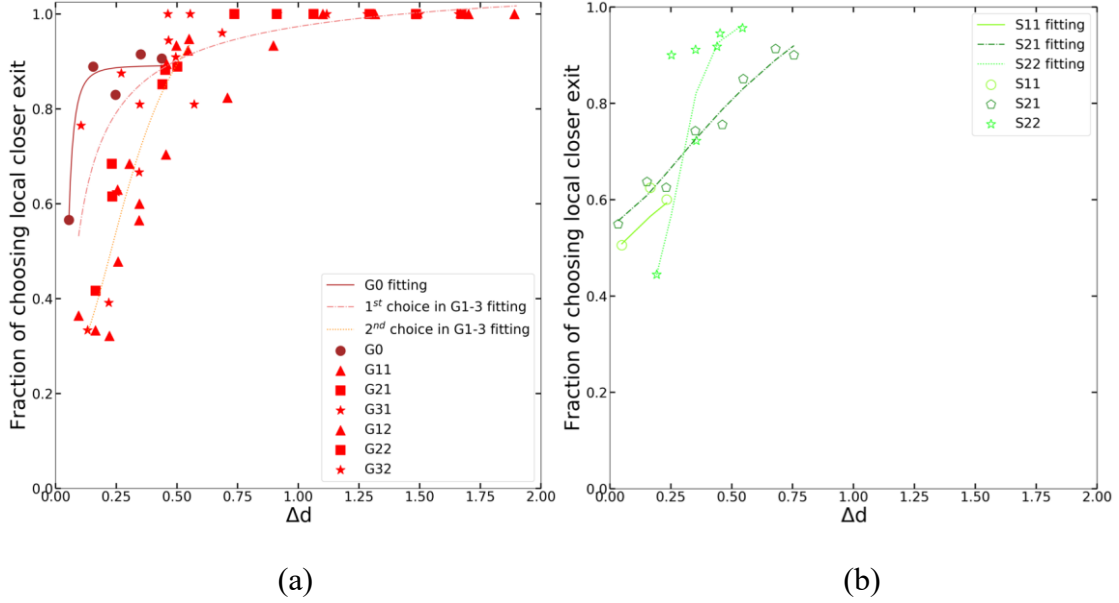

(a)　　　　　　　　　　　　　　(b)

Fig 4. Relation between difference in distance to exits $\Delta d$, and the fraction of pedestrians who choose the local closer exit in the field observation (a) and the virtual experiment (b). G0: the first choice in G0. G11: the first choice in G1. G12: the second choice in G1, and so on. S11: the first choice in S1. S21: the first choice in S2. S22: the second choice in S2, and so on. For computing fractions $p$ and plotting, $\Delta d$ is divided into bins of 0.1m and data points shown are thus based on different numbers of observations. To guide the eye, we fit lines according to $p = A_2 + (A_1 - A_2)/(1 + (\Delta d/\Delta d_0)^a)$ to this data (where $A_1$, $A_2$, $a$ and $\Delta d_0$ are parameters). For a statistical analysis of our data, see section 3.4.

## 3. 2 Density in measurement areas

In a similar way to investigating the preference of pedestrians for nearer exits, we also assess the effect of how well-used exits are. Previous work suggests that pedestrians avoid queues (Bode et al., 2014, 2015b; Bode and Codling, 2013; Liao et al., 2017) and we therefore expect that individuals select exits that are associated with a lower local density of pedestrians. We measure the density difference ($\Delta\rho$) across two neighboring exits as the number of pedestrians per square meter within the measurement areas indicated in Fig. 1 at the same time point as when we also measure the distance $d$ of individuals to exits (see above). We use a method based on Voronoi Diagrams to calculate the density inside measurement areas (Steffen and Seyfried, 2010).

Analogously to our investigation above, for consistence, we calculate $\Delta\rho$ as the difference between the higher and the lower density at two neighboring exits and consider the relationship of this measure with the fraction pedestrians who choose the exit associated with the lower density (Fig. 5). The relationship we find is less clear than for $\Delta d$, but it seems that the fraction of choosing the exit with lower density increases slightly as $\Delta\rho$ increases for the field observations and virtual experiments. A formal statistical assessment of this effect can be found in section 3.4.

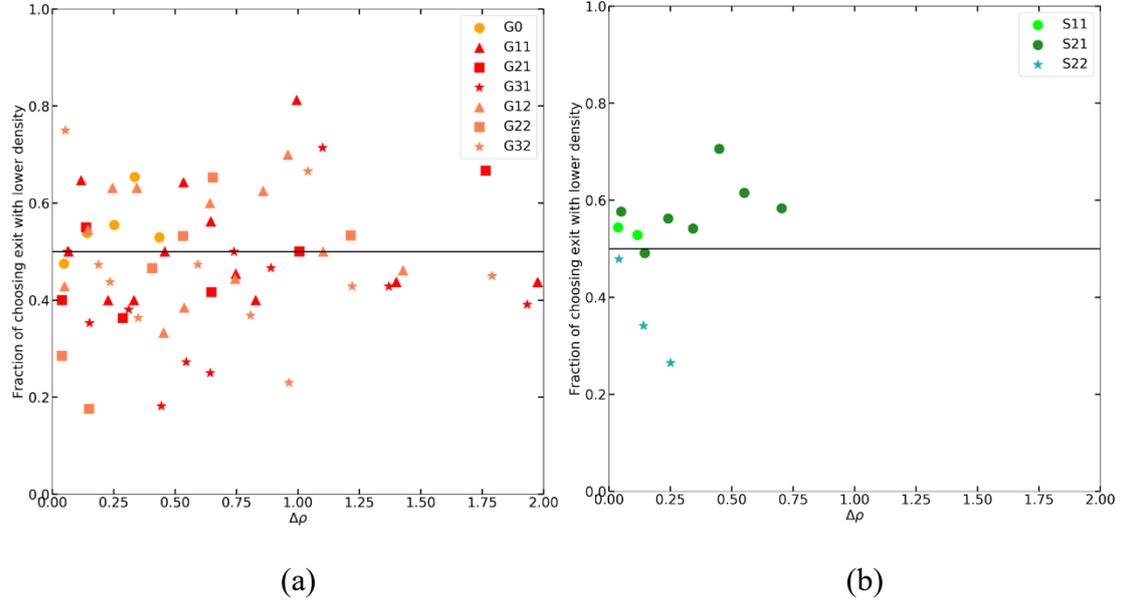

(a) (b)

Fig 5. Relation between difference in density at exits, $\Delta\rho$, and the fraction of pedestrians who choose the less-used exit in the field observation (a) and the virtual experiment (b). G0: the first choice in G0. G11: the first choice in G1. G12: the second choice in G1, and so on. S11: the first choice in S1. S21: the first choice in S2. S22: the second choice in S2, and so on. For computing

fractions, $\Delta\rho$ is divided into bins of 0.1 ped/m² and data points shown are thus based on different numbers of observations. For a statistical analysis of our data, see section 3.4.

## 3. 3 Speed in measurement areas

The final factor we consider is the speed at which pedestrians using exits move. Previous work suggests that pedestrians use time-dependent information, such as the speed at which queues at exits move when making their choices (Bode et al., 2015b). Thus, we might expect that pedestrians might choose exits at which the average speed of pedestrians is higher. We measure the pedestrian speed difference, $\Delta v$, across two neighboring exits as the difference in mean speed of pedestrians inside the measurement areas shown in Fig. 1 at the same time points as for the analysis above. We make use of a method based on Voronoi Diagrams (Steffen and Seyfried, 2010) to calculate the average speeds inside the measurement areas and for consistence, we always subtract the lower from the higher average speed when computing $\Delta v$. The relationship between $\Delta v$ and the fraction of pedestrians who choose the exit associated with the higher average pedestrian walking speed can be seen in Fig. 6. It seems that the fraction of choosing the exit with higher speed increases slightly as $\Delta v$ increases for the virtual experiments while $\Delta v$ has no substantial influence on exit choice in the field observations. A formal statistical assessment of this effect can be found in section 3.4.

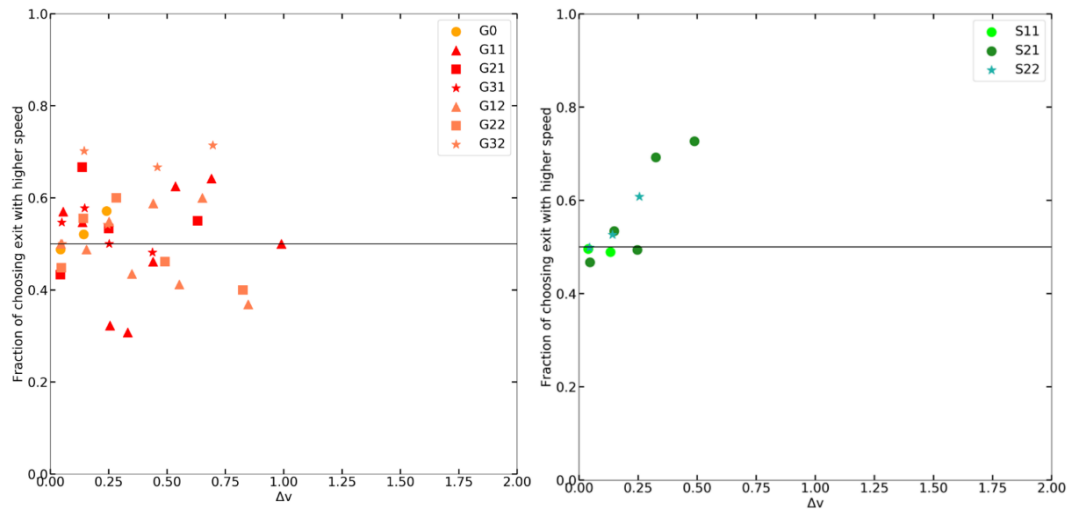

(a) (b)

Fig 6. Relation between difference in average walking speeds at exits, $\Delta v$, and the fraction of pedestrians who choose the exit associated with higher average walking speeds in the field observation (a) and the virtual experiment (b). G0: the first choice in G0. G11: the first choice in G1. G12: the second choice in G1, and so on. S11: the first choice in S1. S21: the first choice in S2. S22: the second choice in S2, and so on. For computing fractions, $\Delta v$ is divided into bins of 0.1 m/s and data points shown are thus based on different numbers of observations. For a statistical analysis of our data, see section 3.4.

## 3.4 Statistical analysis

In this section we formally assess the trends indicated in our preliminary analysis above.

In our field observations and virtual experiment, individuals have the choice between either two or three routes (second choice in G1-G3 and S2). Based on the observation that in the case when individuals have three options almost all pedestrians use one of the two closer exits (except 1 person in G11-G32 and 2 people in S2), we simplify our statistical analysis by only considering a binary choice and do not include data from individuals who opted for the exit furthest away in this analysis. For each individual and each time a route has to be chosen, we record if he or she chooses the exit closer to the top of the page when the scenario is plotted in two dimensions, as shown in Fig 1 and 2. For example, in Fig 1(a), Exit A is the exit closer to the top of the page. This binary variable allows us to record individuals' choices consistently with respect to the three factors introduced above, regardless of whether there are only two options or whether we only consider two out of three options.

We refer to the proportion of individuals who choose the route closer to the top of the page as $P(top\ exit)$ and use logistic regression to investigate the effect of three main explanatory variables, based on the factors introduced above, on the binary variable we record. First, we consider the difference in distance, *d_dist*, to the two exits by subtracting the distance from the location of the individual to the middle of the exit opening from the same measure for the top route. Note that *d_dist* differs from *Δd* introduced above. For example, if the top exit is closer, then *d_dist* is negative and vice-versa. We record this measure at the same time point as used before in sections 3.1.-3.3. In a similar way to the difference in distance between the two exits (*d_dens*), we record the difference in average density (*d_dens*) and walking speeds (*d_speed*) in the measurement areas as additional explanatory variables.

We consider data from the first time and the second time individuals make a choice as independent. To investigate general differences between first and second choice times and differences across observational or experimental scenarios, we additionally consider interactions between the three main explanatory variables (*d_dist, d_dens, d_speed*) and a categorical variable that distinguishes between experimental scenarios, as well as the first time and second time individuals make a choice (i.e. for the observational data, the variable has 7 levels, 'G0', 'G11', 'G12', 'G21', 'G22', 'G31', and 'G32' and for the virtual experiment it has 3 levels, 'S1', 'S21', and 'S22').

We use Likelihood-ratio tests (LR tests) to determine if interaction terms should be included in our statistical analysis. We perform separate tests for interactions of each of the three main explanatory variables with the categorical variable defined above. Thus, for the field observations, in each LR test, we compare a model with a linear predictor comprising an intercept, three main explanatory variables and six interaction terms between experiment type and one of the main explanatory variables with a nested model that does not include the interaction terms (for data from the virtual experiments, we only have two interaction terms).

In the virtual experiment, each participant repeats experimental settings S1 and S2 six times with different initial conditions. This means, we obtain 18 data points from each participant (participants make six choice in S1, and six first and six second choices in S2). Considering the simplicity of the experiment and the similarity of the tasks, it is important to investigate if the behavior of participants changes over time. As habituation or learning effects do not affect our main results, we only report this additional analysis in Appendix C. We use the R programming environment, version 3.2.2, for all statistical analysis (R Core Team, 2013).

Performing this analysis on our field observation data, we find that only interaction terms of the difference in distance between the top and bottom exit with experimental setting substantially improve our statistical model (LR test, $X^2(6)=35.72, p=3.12x10^{-6}$), whereas interaction terms for differences in density (LR test, $X^2(6)=5.44, p=0.49$) or average speed (LR test, $X^2(6)=9.52, p=0.15$) do not. The model fitting results are summarized in Table 1.

Table 1. Statistical analysis of field observation data.

| Explanatory factor | Parameter estimate ± s.e. | Z value | Pr(>|z|) |
|---|---|---|---|
| Intercept | -0.15±0.09 | -1.73 | 0.08 |
| d_dist | -6.81±0.84 | -8.07 | $7.24\times10^{-16}$ |
| d_dens | -0.40±0.14 | -2.90 | $3.78\times10^{-3}$ |
| d_speed | 0.021±0.28 | 0.075 | 0.94 |
| d_dist:G11 | 3.37±0.97 | 3.49 | $4.79\times10^{-4}$ |
| d_dist:G12 | 4.52±0.94 | 4.78 | $1.74\times10^{-6}$ |
| d_dist:G21 | 0.92±1.88 | 0.49 | 0.62 |
| d_dist:G22 | 3.34±1.14 | 2.92 | $3.47\times10^{-3}$ |
| d_dist:G31 | 0.81±1.56 | 0.52 | 0.61 |
| d_dist:G32 | 2.83±1.02 | 2.78 | $5.40\times10^{-3}$ |

We use logistic regression to investigate the effect of the following factors on whether individuals choose the top exit: *d_dist*: distance difference between top exit and bottom exit. *d_dens*: density difference between top exit and bottom exit. *d_speed*: speed difference between top exit and bottom exit. *d_dist:GXX* refers to the interaction terms between experiment type and difference in distance. G11: the first choice in G1. G12: the second choice in G1, and so on. Interaction terms describe the change in the effect *d_dist* has relative to the baseline G0. Test statistics and p-values are for the Wald test, assessing the null hypothesis that the corresponding parameter estimate is equal to zero.

These results suggest that both the difference in distance and the difference in density between exits influence the route choice around obstacles of individuals with the effect of the difference in distance being substantially stronger (see parameter estimates in Table 1). The negative parameter estimates for the difference in distance and density indicate that on average individuals avoid exits further away and exits with a higher density of other people (see also Fig 7). There is no evidence suggesting that the difference in speed has an effect and while the estimate for the intercept suggests that overall there is a small bias towards the bottom exit across experiments, we cannot rule out that this is coincidental (*p=0.08*).

The interaction terms show changes in the effect the difference in distance has relative to the baseline of experiment G0. The estimates of all interaction term parameters are positive, suggesting that the difference in distance has a reduced effect size in all experiments compared to G0. This reduction in effect size is unlikely to have arisen by chance for experiment G11 and, interestingly, for all second choices (G12, G22, G32; all p-values *<0.05*). Fig 7 (c) shows that the transition from high to low $P(top\ exit)$ with increasing *d_dist* is slower for these experiments and second choices. Experiment G11 has the shortest distance $L_2$ from the entrance into the experimental arena to the first set of exits (see Fig. 1 (b)). Thus, these findings could suggest that *d_dist* has a smaller effect if individuals have less time (space, $L_2$) to consider these differences and a similar argument could be made about the more confined situation of the second choice. Fig. 7 (d) shows the effect of difference in densities across exits.

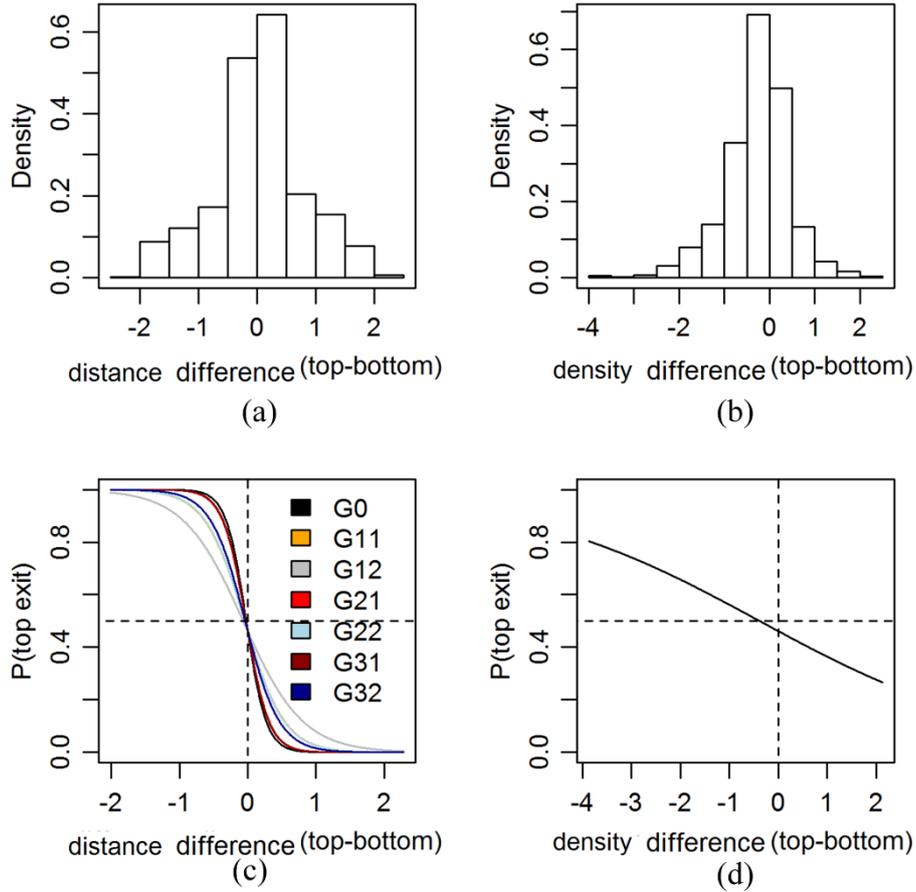

Fig. 7 Observational data: distributions of observed differences between exits (a, b) and model fits to data (c, d). The fraction of choosing the top exit, $P(top\ exit)$, shows a downward trend as the difference in distance, *d_dist,* between the top and bottom exit increases. The difference in density, *d_dens,* shows a qualitatively similar effect. Plots show the model fits over the observed range of the explanatory variables. All explanatory variables that are not varied in plots are set to zero to obtain model fits. The intersection of the dashed lines in (c, d) indicates unbiased choices.

We adopt the same approach for our statistical analysis of the data from our virtual experiment. In contrast to the observational data, Likelihood-ratio tests indicate that interaction terms between both the difference in distance and experimental setting (LR test, $X^2(2)=60.11$, $p=8.52 \times 10^{-14}$) and between the difference in density and experimental setting should be included in our model (LR test, $X^2(2)=11.57$, $p=3.08 \times 10^{-3}$). Interactions between average walking speed and experimental setting do not substantially improve model fit (LR test, $X^2(2)=0.79$, $p=0.67$). The model fitting results are summarized in Table 2.

Table 2. Statistical analysis of virtual experiment data.

| Explanatory factor | Parameter estimate ± s.e. | Z value | Pr(>|z|) |
|---|---|---|---|
| Intercept | -0.14±0.05 | -2.57 | 0.01 |
| d_dist | -2.58±0.48 | -5.37 | $7.77 \times 10^{-8}$ |
| d_dens | -2.99±1.44 | -2.08 | 0.04 |
| d_speed | 0.41±0.42 | 0.99 | 0.32 |
| d_dist:S21 | -0.53±0.53 | -0.99 | 0.32 |
| d_dist:S22 | -2.95±0.57 | -5.15 | $2.64 \times 10^{-7}$ |
| d_dens:S21 | 1.43±1.48 | 0.97 | 0.33 |
| d_dens:S22 | 2.88±1.90 | 1.51 | 0.13 |

We use logistic regression to investigate the effect of the following factors on whether individuals choose the top exit: *d_dist*: distance difference between top exit and bottom exit. *d_dens*: density difference between top exit and bottom exit. *d_speed*: speed difference between top exit and bottom exit. *d_dist:SXX* refers to the interaction terms between experiment type and difference in distance. *d_dens:SXX* refers to the interaction terms between experiment type and difference in density. Interaction terms describe the change in the effect of *d_dist* or *d_dens* relative to the baseline S1. Test statistics and p-values are for the Wald test, assessing the null hypothesis that the corresponding parameter estimate is equal to zero.

These results suggest that overall there is a small bias towards the bottom exit across experiments (intercept), but there is no evidence suggesting that the difference in speeds has an effect. The difference in distance and density between exits influences the route choice around obstacles of individuals and the negative parameter estimates indicate that on average individuals avoid exits further away and exits with a higher density of other people, analogously to the field study data (see also Fig 8). Due to the presence of interaction terms, the p-values and parameter estimates for the main effects of difference in distance and density or the corresponding interaction terms should not be considered in isolation.

The interaction terms between the difference in distance and experimental setting are negative, suggesting that the difference in distance has an increased effect size in all experiments compared to S1. Fig 8 (c) shows that the transition from high to low $P(top\ exit)$ is faster for experimental settings S21 and S22.

Considering the interaction terms between the difference in density and experimental setting, our results suggest that in both S21 and S22 any effect of the difference in density is reduced, and for S22 essentially negligible when compared to the effect the difference in density has in S1 (Fig. 8 (d)).

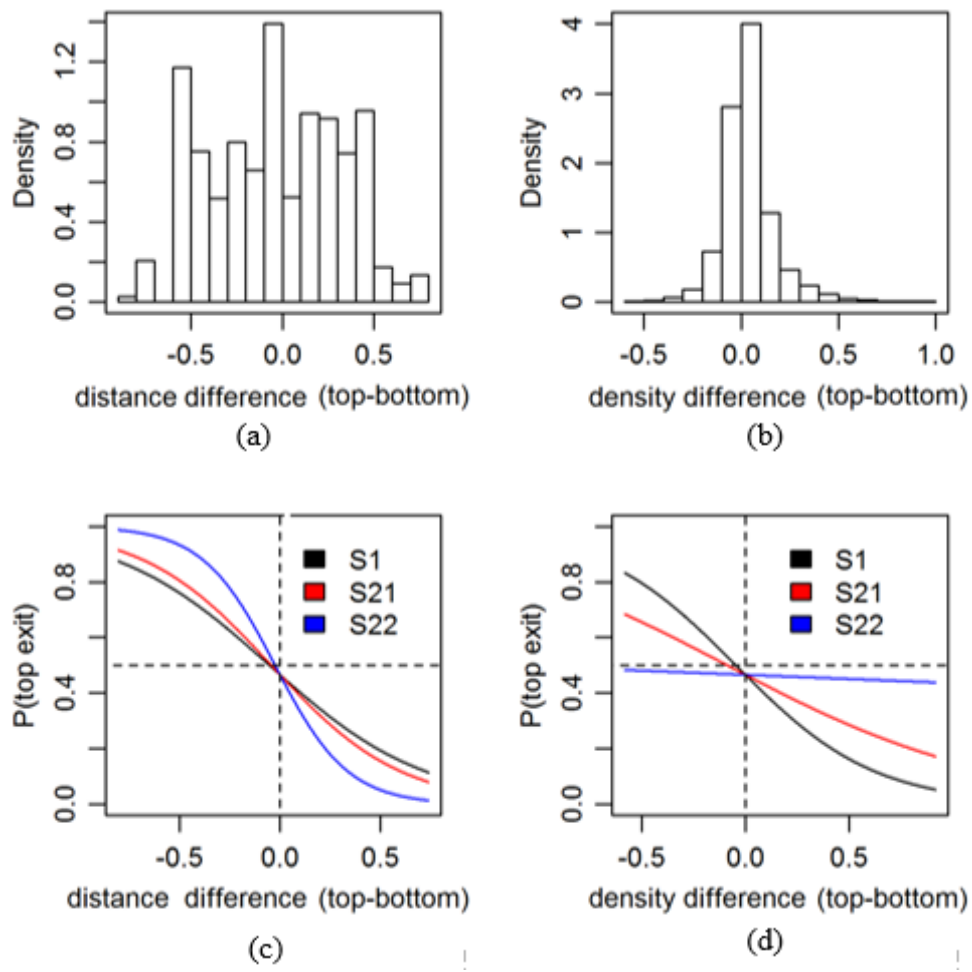

Fig. 8 Virtual experiment data: distributions of observed differences between exits (a, b) and model fits to data (c, d). Panel (c) shows that the fraction of choosing the top exit, $P(top\ exit)$, decreases

as the difference in distance between the top and bottom exit increases. Considering panel (d), in S1, $P(top\ exit)$ decreases with increasing difference in density between top exit and bottom exit, while in S21 and S22, this effect is reduced. S1: the first choice in S1. S21: the first choice in S2. S22: the second choice in S2, and so on. Plots show the model fit over the observed range of the explanatory variables. All explanatory variables that are not varied in plots are set to zero to obtain model fits. The intersection of the dashed lines in (c, d) indicates unbiased choices.

To examine how well the statistical models presented in tables 1 and 2 capture our data and therefore to assess the extent to which the factors we consider explain the route choice around obstacles of pedestrians, we use the logistic regression models as binary classifiers to predict exit selection. Using our models, we can predict $P(top\ exit)$ for each choice in our data. We set a threshold for $P(top\ exit)$, above which we predict individuals to choose the top exit and vice-versa. We then compare these predictions to the observed choices and determine threshold levels that maximize the number of correct predictions (thresholds for $P(top\ exit)$ are 0.61 and 0.53 for field observations and virtual experiments, respectively). Our models correctly predict 82% and 74% of all choices for the field observation and the virtual experiment, respectively. This suggests that we can explain over 74% of exit selections in our data based on distances to exits, densities at exits, experimental conditions and speeds at exits (although speeds are unlikely to aid prediction, based on our statistical analysis).

In summary, we find qualitatively comparable results for our field observation and our virtual experiment. For both data sets, we find evidence for individuals choosing closer exits and avoiding busier exits (i.e. exits associated with a higher density of pedestrians). In neither data set the average speed of pedestrians using an exit appears to influence individuals' choices. To further corroborate these findings, we also compare the data sets by fitting the same statistical models to them (Appendix D). Differences between our findings for the two data sets occur when we consider how the main effects are modulated in different experimental settings. For example, the effect of differences in density across exits appears not to be affected by the experimental settings in the field study, but this density-related effect is reduced in setting S2 (S21 and S22) in the virtual experiment. Moreover, the effect of the difference in distance to two exits is reduced for second choices in the field study, but generally stronger for first and second choices in experimental setting S2 in the virtual experiment. It is possible that some of these detailed differences could be due to differences between the field study and virtual experiment setup. For example, the area in front of the first exit for which we record data is larger in the virtual experiment (see Figs. 1 and 2) and we do not include social ties (e.g. families or groups of friends, see Table 1) into our virtual experiment that may bias individuals' route choice (Bode et al., 2015a). Based on this, we suggest that despite differences in detailed behavioral responses to experimental setting, our simple virtual experiment elicits the predominant route choice around obstacles mechanisms in humans.

Tables 1 and 2 show that the parameter estimates differ between field observations and our virtual experiment. It is important to note that it does not make sense to compare these parameter estimates quantitatively for the following reasons. First, the structure of the models we fit differs across setting (see interactions terms). Second, we have no reason to expect that people will interpret distances, densities etc. in the same quantitative way in the real world and in our virtual experiments (e.g. we cannot expect that 1m difference in distance to exits will have the same effect on pedestrians' route choice in the two settings).

## 4. Discussion

The two main contributions of our work are an investigation of the factors influencing pedestrian route choice behavior around obstacles in the travel path and a direct comparison between this behavior in a field observation and a virtual experiment.

Considering our first contribution, our findings lend additional support to previously suggested route choice mechanisms. We show that in the simple scenarios we investigate, over 74% of choices can be predicted by considering the distance to exits and the number of pedestrians using exits. We find average speeds of pedestrians around exits do not substantially contribute to choices. As suggested in the literature, we find that pedestrians prefer shorter routes and less busy routes (Bode et al., 2014, 2015b; Bode and Codling, 2013; Borgers and Timmermans, 1986; Haghani and Sarvi, 2016; Kemloh Wagoum et al., 2012; Liao et al., 2017; Srinivasan et al., 2017). From the highly simplified scenarios we consider, it is difficult to determine if pedestrians choose the globally or locally shortest route, but our results indicate that individuals prefer the locally closest exit (see Fig. 3). We also note that our results derive from low-pressure experiments and that route choice behavior may change under increased levels of stress (Bode et al., 2015b; Bode and Codling, 2013). One explanation for the interactions between the difference in distance to exits and the experimental setting we find (e.g. the effect of the difference in distance is stronger in S2 compared to S1), could be the differences in total walking length to exits in the different scenarios. For example, rather than considering absolute values of differences in distance to exits, as we do in our analysis, pedestrians may only consider differences in distance that are a sufficiently large proportion of the total walking distance somewhat analogously to models in car following that assume drivers do not distinguish small speed differences (Wiedemann, 1974).

The second contribution of our work is an empirical validation of a simple virtual experiment paradigm for pedestrian route choice behavior around obstacles in the travel path. Our validation approach is based on conducting almost identical studies in the field and in a virtual experiment. If the studies produce comparable results, this indicates that virtual experiments elicit similar behavior in humans to what we observe in real life. We suggest that this type of direct test of the validity of virtual experiment frameworks is important, as otherwise it is not clear to what extent the findings from this work apply in real life (Bode et al., 2015b; Kinateder et al., 2014c). While we find some differences in how experimental settings modulate the effect of the distance to exits and the density at exits, our findings on the main effects of the three factors we study are qualitatively consistent across our field observation and virtual experiment, despite the fact that the participants for our virtual experiment were predominantly university students, whereas the event at which the observational data was recorded attracted a wider range of visitors, including pupils, their parents and students. This is particularly interesting since our virtual experiment presents a highly abstracted setting in which participants have a top-down view and use simple controls to steer their character. Based on this, it may not be necessary to always use highly immersive virtual reality settings for this type of research.

Nevertheless, great care is warranted when extrapolating findings from our, or indeed any, virtual experiment to real world contexts (Kinateder et al., 2014c). Virtual experiments do not faithfully incorporate many of the visual, auditory, olfactory and somatosensory inputs humans experience in real life and therefore any unverified stimulus or context in virtual environments can elicit unintended behavioral responses. Virtual settings could also detach individuals from social norms and cause overly competitive behavior, for example. In our experiments, we note that even though we instruct participants to behave as if they were leaving an exhibition, many of them nevertheless behave competitively and try to exit as quickly as possible during the experiment. Previous work in a pedestrian behavior context suggests similar effects, caused by participants possibly treating experiments as games (Bode and Codling, 2018). Another concern about virtual experiments, especially for ones with more complex controls than ours, is the possibility that the skill of participants to interact with the simulation, based on their computer literacy, could influence their behavior (Bode et al., 2014).

In contrast to experiments using virtual environments, field observations offer the advantage that behavior is observed in a natural setting. Observing pedestrian behavior at events also removes the need to recruit and pay volunteers, making this approach to data collection cost effective. However, data from field observations also has limitations that are important to consider. It is more difficult to control observational settings. This can make data collection more difficult. Some pedestrians in our observations carry umbrellas which means they and sometimes other pedestrians walking nearby are occluded from direct view. This results in errors from the tracking software, which we need to correct manually. Lack of control over observational settings also means that findings from this type of data may be specific to the setting investigated. For example, the composition of the pedestrian crowd studied, such as the presence of children or family groups, could influence findings and it is unclear whether the motivational levels, e.g. relating to time pressure, of all pedestrians are the same or how they differ. Most importantly, the limited extent to which the physical environment and pedestrian behavior can be manipulated for experimental purposes in observational settings restricts the insights that can be gained. Specifically, this means most observational research can only uncover correlations in data rather than establishing causal links between variables describing pedestrian behavior.

Virtual experiments have advantages compared to laboratory experiments and field observations when studying behavior of pedestrians, such as route choice (Bode et al., 2015b; Kinateder et al., 2014c). They are cheaper to run, as one individual can interact with a simulated crowd and they can be conducted remotely (e.g. online), they facilitate testing high-pressure scenarios that could be unsafe when performed with groups of volunteers and they allow researchers to fully control the experimental setting and thereby to focus on precise aspects of behavior. Therefore, we hope our work is a useful starting point for more research on developing an understanding of how behavior in virtual experiments relates to real-life contexts and to thereby establish the full potential and appropriate use of available experimental paradigms.

While more work is needed to fully establish the validity of virtual experiments for studying pedestrian behavior, our findings can already be used to inform future work on route choice. First, we suggest that future work focusing on simple route choice scenarios in which the length and crowdedness of routes vary could initially be performed using simple virtual experiments to trial behavioral responses related to additional factors, such as time pressure, individual characteristics or signage (similar to work in (Bode et al., 2014; Bode and Codling, 2013; Haghani and Sarvi, 2016, 2017a)). This could be an efficient and cost-effective approach to select factors for further investigation in more realistic experiments. Second, we suggest that route choice over longer routes involving not one or two choices, as in our experiment, but dozens of times a choice has to be made could also be investigated using virtual experiments, provided the scenario investigated forces pedestrians to make choices locally, rather than using globally visible landmarks, such as tall buildings. Finally, the habituation effects we found (see also Appendix C) point to an alternative use for virtual experiments related to safety management. Rather than further investigating the natural route choice behavior in pedestrians, as suggested in the first point above, the validity of these experiments could be exploited to investigate changes in the behavior of building occupants as they habituate to overly-frequent fire drills, for example. However, as already indicated above, we consider the main contribution of this study to be a first step towards validating virtual experiments as a useful experimental paradigm for studying pedestrian behavior and we expect that further validation efforts will substantially increase the usefulness of virtual experiments beyond what we suggest here.

# Appendix A – information on age and social group composition in the field observation

Table A.1 shows information on the social group composition of pedestrians in the different route choice behavior around obstacles scenarios in our field observation. We infer this information, as well as the gender of pedestrians manually from the video recordings. We identify pedestrians walking closely together and interacting via gestures and head turn towards each other as belonging to the same social group. The gender of pedestrians is identified from their clothes and hairstyle.

Several studies show that gender and social groups can impact pedestrian movements. For example, Faria et al. (Faria et al., 2010) point out that genders appear to play an important role in decisions of when to move when crossing a road. In addition, the size of social groups has been shown to affect walking speed in medium density situations (Bandini et al., 2012). To enable future uses of our data for this type of research and because we cannot publish the original video recordings for ethical reasons (individuals may be identifiable), we record information on gender and social groups.

We deliberately do not use this additional data in the present study, as we focus on comparing a simple scenario between observational data and a virtual experiment without the added complexity of implementing gender differences and social groups in our virtual environment.

Our method of inferring gender and groups is identical to previous work (Li et al., 2012).

Table A.1 Group composition of field observations.

|     | Single      | Dyads      | Triads     | ≥ Four-person |
|-----|-------------|------------|------------|---------------|
| G0  | 37(12, 25)  | 61(57,70)  | 19(31,26)  | 3(5,7)        |
| G1  | 40(21,19)   | 61(54,68)  | 12(12,24)  | 2(6,8)        |
| G2  | 37(14,23)   | 21(20,22)  | 2(0,6)     | 1(2,2)        |
| G3  | 46(15,31)   | 50(34,66)  | 6(7,11)    | 0             |

Note: G0 refers to the first scenario and G1-3 refers to the second scenario with different settings for $L_1$ and $L_2$. The numbers in the brackets represent the number of females and males respectively. The former was marked with underlines.

# Appendix B – trajectories of pedestrians in the field

observation

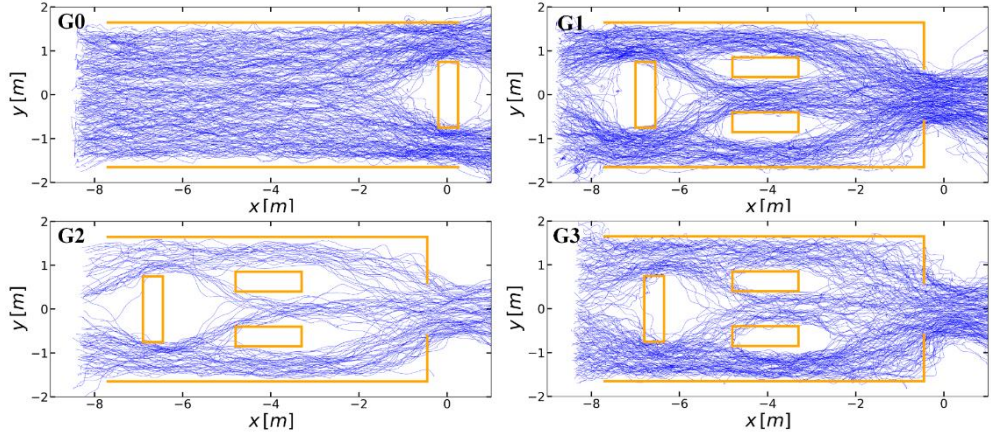

Fig. B.1 Trajectories in the field observation. The orange lines and hollow rectangles indicate barriers. The blue lines are the trajectories of every pedestrian.

Fig B.1 shows the trajectories in the field observation. It can be seen that parts of the trajectories overlap with the obstacles or outside of the geometries. That is because the tables used for constructing the geometries are lower than pedestrian and the head of the pedestrians crosses the border sometimes. In this case, we analyze the trajectories combined with the video recordings in this study to improve the reliability of the results.

## Appendix C – habituation effects in virtual experiment

Each participant in our virtual experiment completes 12 consecutive tasks. Considering the simplicity of the experiment and the similarity of the tasks, it is important to investigate if participant behavior changes with the number of tasks performed. To do so, we include the sequential task number (from *1* to *12*) as an additional explanatory variable in our statistical analysis.

We use a Likelihood-ratio test to compare the model presented in Table C.1 with a model that additionally includes task number and interactions of task number with all other predictors as an explanatory variable. Our results suggest that considering task number improves the model fit (LR test, $X^2(12)=154.59$, $p<2.0\times10^{-16}$ – below the numerical precision threshold of our software). We use further LR tests to assess improvements to model fit separately for each interaction term involving task number (details omitted). Excluding terms that do not improve model fit results in the model presented in Table C.1. We find that all main effects discussed in the main text remain unaffected. Habituation of participants to the tasks appears to have two consequences. First, over time, participants develop a bias for bottom exits after an initial preference for the top exit (intercept and task number effect). Second, participants' preference for closer exits gets stronger over time (interaction between difference in distance between exits and task number).

Table C.1. Statistical analysis of virtual experiment data considering habituation effects.

| Explanatory factor | Parameter estimate ± s.e. | Z value | Pr(>|z|) |
| --- | --- | --- | --- |
| Intercept | 0.71±0.11 | 6.46 | $1.04 \times 10^{-10}$ |
| d_dist | -1.92±0.59 | -3.22 | $1.29 \times 10^{-3}$ |
| d_dens | -3.69±1.49 | -2.48 | 0.01 |
| d_speed | 0.37±0.42 | 0.88 | 0.38 |
| task_no | -0.14±0.02 | -8.64 | $< 2.0 \times 10^{-16}$ |
| d_dist:S21 | -0.28±0.55 | -0.51 | 0.61 |
| d_dist:S22 | -2.68±0.59 | -4.54 | $5.65 \times 10^{-6}$ |
| d_dens:S21 | 2.20±1.52 | 1.45 | 0.15 |
| d_dens:S22 | 3.65±1.95 | 1.87 | 0.06 |
| d_dist:task_no | -0.15±0.06 | -2.60 | $9.35 \times 10^{-3}$ |

We use logistic regression to investigate the effect of the following factors on whether individuals choose the top exit: *d_dist*: distance difference between top exit and bottom exit. *d_dens*: density difference between top exit and bottom exit. *d_speed*: speed difference between top exit and bottom exit. *task_no*: the task number. *d_dist:SXX* refers to the interaction terms between experiment type and difference in distance. *d_dens:SXX* refers to the interaction terms between experiment type and difference in density. Interaction terms describe the change in the effect of *d_dist* or *d_dens* relative to the baseline S1. Test statistics and p-values are for the Wald test, assessing the null hypothesis that the corresponding parameter estimate is equal to zero.

## Appendix D – supplementary statistical analysis

In section 3.4 in the main text, we fit different statistical models to the two different data sets. We performed this analysis to carefully investigate any differences in participant behavior depending on the experimental conditions. To facilitate a more direct comparison between the observational data and the virtual experiment data, we here perform an additional analysis where we fit the same statistical models to both data sets. Due to the different number of experimental conditions, we only focus on the main effects related to distance, density and speed in this analysis. We find good qualitative and even good quantitative agreement in results across the data sets suggesting that individuals preferentially choose closer exits, exits that are used by fewer others and that the speed of individuals using exits does not seem to affect decisions substantially in our data (Tables D.1 and D.2).

Another way to perform this comparison, is to compare the virtual experiment data with the observational data, but to only consider the experimental condition out of G1, G2 and G3 that is the most similar to condition S2 (i.e. G1). Using the same statistical models as shown in Tables D.1 and D.2 we again find a good qualitative match in results for observational and virtual experiment data (not shown).

Table D.1. Supplementary statistical analysis of field observations data.

| Explanatory factor | Parameter estimate ± s.e. | Z value | Pr(>|z|) |
|---|---|---|---|
| Intercept | -0.16±0.09 | -1.86 | 0.063 |
| d_dist | -3.92±0.24 | -16.23 | $<2.0\times10^{-16}$ |
| d_dens | -0.42±0.14 | -3.13 | $1.73\times10^{-3}$ |
| d_speed | 0.056±0.29 | 0.19 | 0.85 |

We use logistic regression to investigate the effect of the following factors on whether individuals choose the top exit: *d_dist*: distance difference between top exit and bottom exit. *d_dens*: density difference between top exit and bottom exit. *d_speed*: speed difference between top exit and bottom exit. *task_no*: the task number. Test statistics and p-values are for the Wald test, assessing the null hypothesis that the corresponding parameter estimate is equal to zero.

Table D.2. Supplementary statistical analysis of virtual experiment data.

| Explanatory factor | Parameter estimate ± s.e. | Z value | Pr(>|z|) |
| --- | --- | --- | --- |
| Intercept | -0.17±0.05 | -3.45 | $5.72 \times 10^{-4}$ |
| d_dist | -4.08±0.17 | -23.95 | $<2.0 \times 10^{-16}$ |
| d_dens | -1.54±0.39 | -3.94 | $8.17 \times 10^{-5}$ |
| d_speed | 0.63±0.42 | 1.50 | 0.14 |

We use logistic regression to investigate the effect of the following factors on whether individuals choose the top exit: *d_dist*: distance difference between top exit and bottom exit. *d_dens*: density difference between top exit and bottom exit. *d_speed*: speed difference between top exit and bottom exit. *task_no*: the task number. Test statistics and p-values are for the Wald test, assessing the null hypothesis that the corresponding parameter estimate is equal to zero.

# Acknowledgements

This work was supported by the National Key Research and Development Program of China (Grant No. 2018YFC0808600), the National Natural Science Foundation of China (Grant No. 71704168), from Anhui Provincial Natural Science Foundation (Grant No. 1808085MG217) and the Fundamental Research Funds for the Central Universities (Grant No. WK2320000040).